# Ternary nickel hydrides: a platform for unconventional superconductivity and quantum magnetism

Mateusz Domański[1]*, Antonio Santacesaria[2], Paolo Barone[3], José Lorenzana[2]* and Wojciech Grochala[1]

(1) Center of New Technologies, University of Warsaw, Żwirki i Wigury 93, 02-089, Warsaw, Poland
(2) ISC-CNR Institute for Complex Systems and Department of Physics, Sapienza University, Piazzale Aldo Moro 5, 00185, Rome, Italy
(3) SPIN-CNR Institute for Superconducting and other Innovative Materials and Devices, Area della Ricerca di Tor Vergata, Via del Fosso del Cavaliere 100, 00133 Rome, Italy
*E-mail: m.domanski@cent.uw.edu.pl, jose.lorenzana@cnr.it

**Abstract.** *Hydrides are famous for the possibility of reaching high-$T_c$ superconductivity under high pressure within a conventional mechanism. Here we propose ternary nickel hydrides $MNiH_2$ (M = Li, Na) as new materials that mirror key aspects of unconventional superconducting cuprates and nickelates while presenting unique characteristics. Compared to Ni oxides, Ni-H bands are wider due to shorter bond lengths and show a smaller charge transfer energy. This leads to a larger scale of magnetic interactions than for $LaNiO_2$, which previous works in cuprates suggest should lead to a larger $T_c$. The presence of an interstitial anionic electron orbital, which hybridizes with the $d_{z^2}$ band, induces self-doping in the Ni $d_{x^2-y^2}$ band, enabling metallicity and superconductivity in stoichiometric forms. A t-J model computation yields dominant $d_{x^2-y^2}$ superconducting symmetry for all doping levels. This, combined with the improved thermodynamic stability over known $Ni^{1+}$ oxides, positions $MNiH_2$ as compelling materials for unconventional high-$T_c$ superconductivity under standard conditions.*

*Introduction*. Copper oxide materials exhibit the highest superconducting critical temperatures observed at standard pressures. Since their discovery over 35 years ago[1] numerous systems analogous to cuprates have drawn attention[2–5]. Notably, research in layered nickelates has recently led to obtaining a thin-layer hole-doped neodymium nickel(I) oxide $Nd_{0.8}Sr_{0.2}NiO_2$ which shows superconductivity with an onset $T_c$ of 15 K[6]. It is the first case when a material containing $d^9$ cations other than $Cu^{2+}$ shows superconductivity.

In the case of cuprates, strong antiferromagnetic exchange coupling $J$, small charge-transfer energy ($\Delta_{CT}$), together with high anion-cation hopping are considered properties that enable superconducting pairing *via* the spin-fluctuation mechanism[7–10]. Despite the analogous $d^9$ electronic configuration, there are differences between $Ni^{1+}$ and $Cu^{2+}$ oxide systems. Due to low cation charge, $Ni^{1+}$ $d_{x^2-y^2}$ levels have considerably higher energy compared to $Cu^{2+}$. This means that $\Delta_{CT}$ in nickelates is relatively larger than cuprates as confirmed by detailed computations[4,11–13]. Therefore, nickelates are near the Mott–Hubbard regime instead of the charge-transfer case common in cuprates[12–14]. Nevertheless, nickelates are covalent and both theoretical[15] and experimental[16] estimates of $J$, though smaller, are of the same order as in cuprates. Despite that, it has been difficult to detect magnetic order in $LaNiO_2$ or $NdNiO_2$[17–19] which may result from defects due to the poor thermodynamic stability of $d^9$-nickelates.

In this paper, we introduce Ni *hydrides* as a new family of cuprate analogs. Cuprate-like materials are useful if they have similarities with their siblings but are not identical because they can potentially reveal the factors determining high-$T_c$ superconductivity and provide new materials for applications. Furthermore, spin-½ systems are interesting also for the possibility of hosting exotic magnetic phases[20]. Though hydrides are often considered conventional BCS superconductors, *i.e.,* benefitting from high electron-phonon coupling, here we use density functional theory (DFT) to survey $Ni^{1+}$ hydrides primarily as a $d^9$ system exhibiting magnetic effects which are typical precursors of unconventional superconductivity. Already known, fcc nickel hydride $NiH_{1-x}$ shows Pauli paramagnetism for $0 < x \leq 0.3$ with an enhanced susceptibility due to correlations[21]. Unfortunately, spontaneous H desorption makes studying this material challenging[21–23].

Herein, we propose $MNiH_2$ (M = Li, Na) with a $CaCuO_2$ type crystal structure (CCO) as two new cuprates analogs with remarkable properties. Besides analyzing the effects of oxide to hydride substitution and the electronic parameters, we discuss the thermodynamic stability of the proposed hydride compounds.

We notice several important differences upon changing from oxides to hydrides. Hydrogen electron affinity is quite low (0.75 eV vs 1.46 eV for oxygen)[24] and Madelung energies will be substantially different for monovalent ions and short bond lengths in hydrides compared to the divalent oxides. All these factors are expected to have a large impact on $\Delta_{CT}$. Furthermore, $T_c$ in cuprates[25] is believed to correlate with the splitting energy of the two $e_g$ orbitals, $\Delta_{e_g}$. In the hydrides, a shorter H-Ni bond and the lower charge of the anions can have a strong impact on $\Delta_{e_g}$ which needs to be assessed.

*Computations.* Electronic structure calculations were performed using Density Functional Theory (DFT) as implemented in VASP[26] within the projector-augmented-wave approach[27]. We used the generalized gradient approximation (GGA) functional optimized for solids PBEsol[28], setting the k-spacing to 0.15 Å$^{-1}$ and the plane-wave energy cutoff to 520 eV. The electronic density of states (DOS) and band gaps were calculated using the tetrahedron method with Blöchl corrections. To highlight the anionic character of hydrogen in the present context, we follow Refs.[29,30] and increase the Wigner-Seitz H radius to 1.0 Å. Both for the paramagnetic and the magnetic computations below we used the DFT GGA+$U$ method within the formulation Ref.[31]. For the paramagnetic computations, we used $J$ = 1 eV and $U$ = 6 eV which seems optimum for $Ni^{1+}$[12,32,33]. VESTA[34] was used for structure and volumetric properties visualization.

As for superconducting oxides[11,12,35,36], the band structure can be studied in the magnetically ordered state or suppressing magnetism. The latter is useful to estimate the fundamental electronic parameters of the system and is the relevant structure when doping suppresses magnetism and drives the system into a metallic paramagnetic state. Therefore, we start by describing the results in the nonmagnetic state. We consider the electronic structure of $MNiH_2$ in a CCO phase. Here $NiH_2$ layers present in the structure, **Fig. 1(a)**, play the role of $CuO_2$ planes in cuprates. **Figure 1(c)** shows the orbital-resolved DFT band structure and DOS of $LiNiH_2$ highlighting the Ni $d_{x^2-y^2}$, $d_{z^2}$, and H states, and their mixing. We provide similar plots for $NaNiH_2$, $CaCuO_2$, and $LaNiO_2$ in the (**Fig. S1** in Supplemental Material[37]).

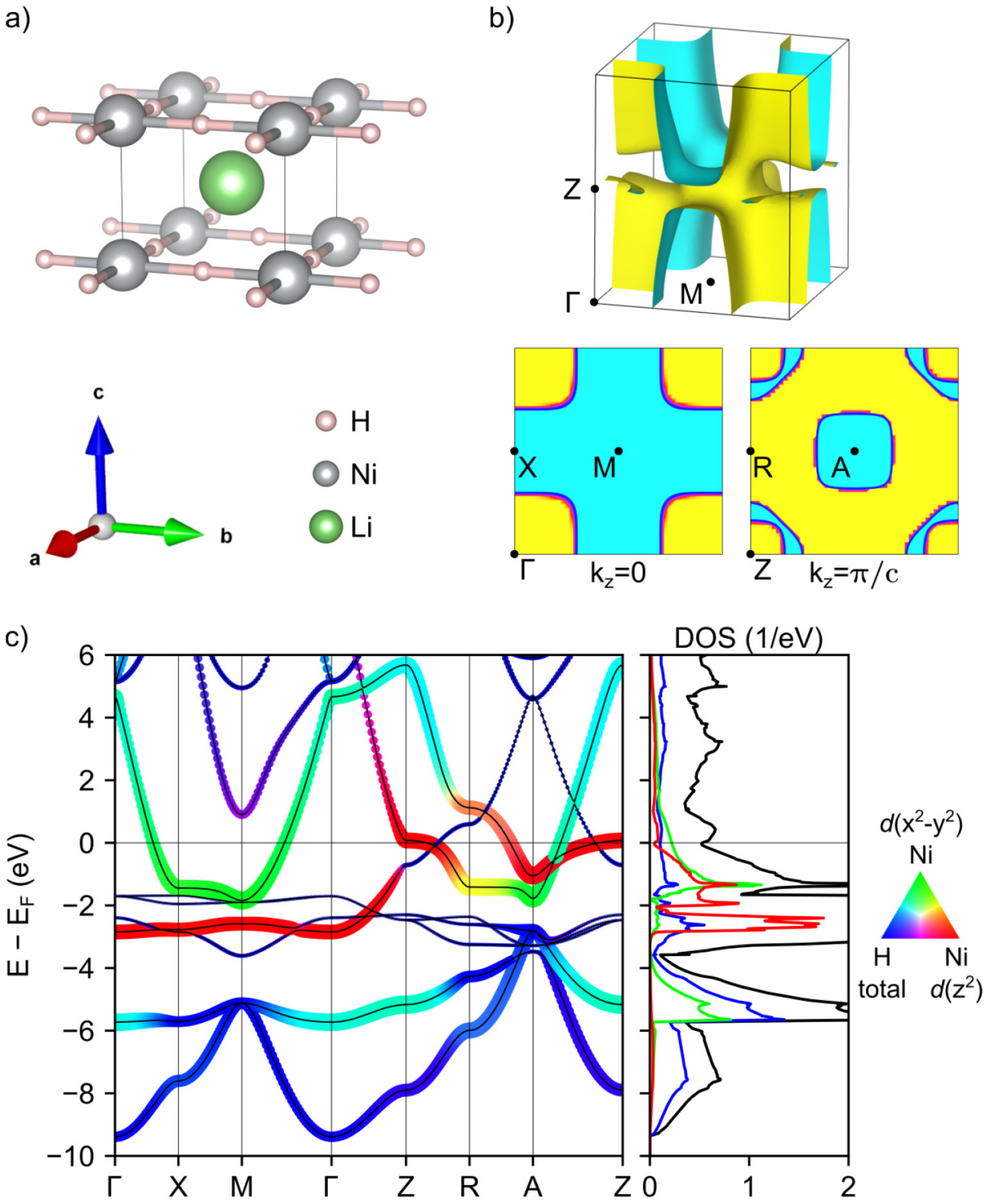

**Fig. 1.** (a) CCO type crystal structure of $LiNiH_2$ hydride. (b) Fermi surface calculated using Wannier interpolation[31,32]. (c) GGA+U band structure and density of states for $LiNiH_2$. Black lines indicate eigenvalues of energy bands while circle diameters indicate the sum of Ni $d_{x2-y2}$, $d_{z2}$, and H *s* states projections in each band. The colors of Maxwell triangle vertices were used to encode the corresponding projected density of states.

The filled H *s* orbitals in $MNiH_2$ play the role of the filled ligand *p* orbitals in oxides. The even parity of the ligand orbitals leads to strong (vanishing) mixing of Ni $d_{x^2-y^2}$ and the ligand at the $\Gamma$-point ($\mathrm{M}$-point) in the hydrides, i.e. exactly the opposite to what occurs in oxides with odd parity ligand orbitals. Another difference is that in cuprates the anion non-bonding bands are narrow, and the bonding band is broad. Here the roles are interchanged. Indeed, among the two lowest bands the lower one is nearly pure H *s* while the upper narrower band is the one mixed with the $d_{x^2-y^2}$ orbital, *c.f.* **Fig. 1(c)**. This indicates that ligand-ligand hopping is larger than in cuprates.

Finally, different from the usual situation in cuprates, the Ni $d_{z^2}$ has a strong dispersion along the $\Gamma-\mathrm{Z}$ path in momentum space. Furthermore, the strong mixing of Ni $d_{x^2-y^2}$ states with H *s* states at $\Gamma$ (0, 0, 0) is mimicked at $\mathrm{Z}$ (0, 0, π/*c*) but the band dispersion is quite different depending on adjacent layers being in phase ($\Gamma-\mathrm{X}-\mathrm{M}-\Gamma$) or in antiphase ($\mathrm{Z}-\mathrm{R}-\mathrm{A}-\mathrm{Z}$). For $\Gamma-\mathrm{X}-\mathrm{M}-\Gamma$ there is only one band crossing the Fermi energy, $E_F$, namely a $d_{x^2-y^2}$ antibonding band like in cuprates (hereafter the "$d_{x^2-y^2}$ band"). In contrast, for the $\mathrm{Z}-\mathrm{R}-\mathrm{A}-\mathrm{Z}$ path, three bands intersect with $E_F$.

The band dispersion can be clarified by computing maximally localized Wannier functions (MLWFs)[38] as implemented in WANNIER90[39]. We have found that five Ni *d* orbitals and two H *s* orbitals do not allow to describe all the bands crossing the Fermi level and an additional orbital is needed which localizes in between two nickels along *c*, **Fig. S2(b)**[37]. Notice that in a perovskite structure, this position would be occupied by an H anion. Although such positions are vacant the system generates an interstitial anionic electron (IAE) orbital. Systems with a significant occupation of an IAE orbital are known as electrides and can exhibit remarkable properties including superconductivity[40–43]. The $d_{z^2}$ orbital mixes strongly with the IAE orbital which together with the small *c* lattice constant explains the strong dispersion. Importantly, there is a partial filling of the IAE band which leads to self-doping.

Interstitial orbitals are delocalized leading to a quasi-free electron dispersion which is challenging to capture with maximum localized Wannier orbitals. Still, the Wannier model reproduces well the electronic structure close to the Fermi level and below, **Fig. S2(a)**[37]. **Figure 1(b)** shows the Fermi surface computed from the Wannier band structure for $LiNiH_2$. Provided that $\Gamma$ and $\mathrm{M}$ are exchanged (due to the different parity of the ligands), cuts at constant $k_z$ resemble a cuprate Fermi surface, except for a relatively narrow region around the $Z$ point where the main Fermi surface changes from having a large hole pocket centered at $\Gamma$ to a large electron pocket centered at $\mathrm{A}$ (referred to as "neck") with satellite pockets related to the extra band crossings. More precisely, $LiNiH_2$ shows three Fermi surface sheets which, moving outwards from $\mathrm{A}$, correspond to the crossings of $d_{x^2-y^2}$, IAE, and $d_{z^2}$ bands. In the case of $NaNiH_2$, the nearly flat $d_{z^2}$ band in the $\mathrm{A}-\mathrm{Z}$ trajectory remains below the Fermi level and only two Fermi surface sheets appear for $k_z$=π/c corresponding to $d_{x^2-y^2}$ and IAE bands, **Fig. S1(b)**.

For $NaNiH_2$, integrating the three-dimensional IAE pocket we find 4% of electrons per Ni which translates into 4% of hole doping of the $d_{x^2-y^2}$ band. In cuprates, this is about the critical doping for superconductivity[44]. For $LiNiH_2$, instead, most of the holes go to the $d_{z^2}$ band and the $d_{x^2-y^2}$ band remains almost half-filled (i.e., doping is less than 0.1%).

We summarize on-site energies and hopping integrals obtained from the Wannier analysis in **Table 1**, where we also include results for oxide reference systems. We can use the difference in diagonal energies between $d_{x^2-y^2}$ and ligand orbitals as a proxy to $\Delta_{CT}$. In the context of cuprates, a lower $\Delta_{CT}$ is associated with a higher $T_c$[45]. For $LaNiO_2$ we find 4.2 eV, a value much larger than for $CaCuO_2$ (1.2 eV) indicating that nickelates are much less covalent than cuprates and in accord with previous results[11,12,35,36]. Our value for the cuprate is smaller than in the literature (2.7 eV[12] and 2.0 eV[36]) probably due to differences in the used functional and computational parameters. For both $MNiH_2$ we find 2.2 eV indicating a much covalent bonding than in $LaNiO_2$ and closer to a cuprate.

We also compare $e_g$ energy splitting for oxides and hydrides. The physically relevant quantity is the position of the $e_g$ levels *after* hybridization with the ligands[46]. To estimate this quantity, we follow Refs.[12,47] and compute the band centroid $\overline{\epsilon}(i) = \int g_i(E)EdE \,/ \int g_i(E)dE$ where $g_i(E)$ is the DFT projected density of states for orbital $i$ as shown in the DOS panel of **Fig. 1(c)**. The integral is restricted to the antibonding bands, for example in the case of $LiNiH_2$ we integrate in the interval (−3 eV, 6 eV). For $CaCuO_2$ and $LaNiO_2$ we obtain $\bar{\epsilon}(d_{x^2-y^2}) - \bar{\epsilon}(d_{z^2}) = 2.7$ eV and 1.9 eV respectively, close to the values in Ref.[12]. For $LiNiH_2$ and $NaNiH_2$ the splittings are smaller, 1.3 eV in both cases. This may be detrimental for superconductivity although both superconducting nickelates and cuprates with an active role of $d_{z^2}$ orbitals have been reported[32,48].

**Table 1.** Calculated on-site energies and the nearest-neighbor hopping integrals for $LiNiH_2$, $NaNiH_2$, $CaCuO_2$, and $LaNiO_2$ derived from the MLWF transformation, provided in eV. The ligand orbital refers to the O $p_x/p_y$ orbital parallel to the bond in the case of oxides and H *s* orbital in the case of hydrides, *d* orbitals refer to Ni/Cu sites. We also report the lattice constants used for GGA+U electronic structure calculations of paramagnetic states that have been optimized for collinear magnetic ground states, using $U$ = 6 (Ni), 7 (Cu) eV. Calculated lattice constants of $CaCuO_2$ and $LaNiO_2$ agree with reported experimental values[49,50] within a 2% margin.

| | | $LiNiH_2$ | $NaNiH_2$ | $CaCuO_2$ | $LaNiO_2$ |
|---|---|---|---|---|---|
| Wannier onsite energies (eV) | $\varepsilon(d_{xy})$ | −2.83 | −2.56 | −4.19 | −3.42 |
| | $\varepsilon(d_{xz}/d_{yz})$ | −2.32 | −1.96 | −3.86 | −2.73 |
| | $\varepsilon(d_{x2-y2})$ | −1.62 | −1.37 | −2.43 | −1.21 |
| | $\varepsilon(d_{z2})$ | −0.64 | −1.15 | −3.15 | −2.09 |
| | $\varepsilon$(*vacancy*) | 2.47 | 1.36 | 2.77 | 0.52 |
| | $\varepsilon$(*ligand*) | −3.79 | −3.56 | −3.59 | −5.36 |
| | $\Delta_{eg}$ | −1.0 | −0.2 | 0.7 | 0.9 |
| | $\varepsilon(d_{x2-y2})$ − $\varepsilon$(*ligand*) | 2.2 | 2.2 | 1.2 | 4.2 |
| Wannier hopping integrals (eV) | ligand − $d_{x2-y2}$ | 1.55 | 1.29 | 1.41 | 1.30 |
| | ligand − $d_{z2}$ | 1.22 | 0.77 | 0.64 | 0.29 |
| | H *s* − *s* (O $p_x - p_y$) | 0.79 | 0.77 | 0.52 | 0.61 |
| lattice constants (Å) | *a*, *b* | 3.22 | 3.36 | 3.80 | 3.92 |
| | *c* | 2.74 | 3.34 | 3.14 | 3.30 |

Inspecting hopping integrals in hydrides, one finds ligand-metal and ligand-ligand hoppings larger than in the oxides which imply wider bands and strong covalency. In the case of $NaNiH_2$ the lattice constants are larger, so hopping integrals become smaller than in $LiNiH_2$ making the former more correlated than the latter.

*Minimal model.* We can fit the bands close to the Fermi level with a tight-binding (See SI for details). The one-band model of cuprates here needs additional bands to represent the pockets. For $NaNiH_2$ two bands are enough which do not hybridize significantly among themselves. **Figure S3**[37] shows the band structure and the Fermi surface for $NaNiH_2$, while **Table S1** reports the parameters. This provides a fairly good fit of the bands near the Fermi surface and of the Fermi surface itself [c.f. **Figs. S1(b)** and **S2** of Ref.[37]]. Both the tight-binding fit and the DFT results have a weak $k_z$ dependence of the Fermi surface until the neck where it changes rapidly. Therefore, for a sizable portion of the Brillouin zone (excluding the neck), the Fermi surface can be characterized by the nearest neighbor, *t*, and next-nearest neighbor, *t'*(0) effective planar hopping parameters. We find |*t*|=0.70 eV considerably larger than the value for cuprates (|*t*|=0.43 eV) indicating wider bands and a weaker correlation regime. The hopping ratio, which in cuprates has been shown to correlate with $T_c$, here takes the value *t'(0)/t=0.43* and is like cuprates with the largest critical temperature.[51] Notice that both *t* and *t'* have the same sign here while the sign is different in cuprates. However, a shift of the Brillouin zone zone $(k_x, k_y, k_z) \rightarrow \left(k_x + \frac{\pi}{a}, k_y + \frac{\pi}{a}, k_z\right)$ flips the sign of *t* keeping the sign of *t'*. Since this shift maps the hydride band structure into a cuprate band structure the two signs compensate and the DOS has the same form as for a cuprate, i.e., the van Hove singularity is found for hole doping.

*Magnetism*. Since in heavy-fermions[52], iron-based superconductors[53], and cuprates[54] superconductivity occurs in proximity to magnetically ordered phases and near a quantum critical point, we investigated Ni hydrides magnetism. Because of self-doping incommensurate magnetic order is possible[55]. Therefore, we used the generalized Bloch condition to calculate the energy of spin spirals with an arbitrary pitch. We set the onsite Hubbard term $U$ = 7 eV for *d* orbitals, as used in the literature for Cu [56,57] and Hund's rule $J$ = 1 eV. We intentionally overestimated $U$ correction for Ni also to equal 7 eV to ease the comparison with Cu (**Fig. 2**).

In the case of $LiNiH_2$ the ground state is close to G-type antiferromagnetic with a small shift, in the ordering wave-vector (q = π/*a*, π/*a*, 0.88π/*c*). However, the commensurate G-type antiferromagnetic (q = π/a, π/a, π/c) is almost degenerate with an energy increase of just +1.5 meV/Ni. This, together with an almost degenerate C-type antiferromagnetic state (+2.2 meV/Ni), indicates a weak antiferromagnetic coupling along *c*, i.e. quasi two-dimensional magnetism as in the cuprates.

Interestingly, for $NaNiH_2$ the ground state is a spin-spiral with three possible ordering wavevectors (⅔π/*a*, ⅔π/*a*, 0), (⅔π/*a*, ⅔π/*a*, π/*c*) and (⅔π/*a*, ⅔π/*a*, ⅔π/*c*) almost degenerate in energy. This is in accord with the fact that, at the mean-field level, doping an antiferromagnetic insulator tends to produce spiral phases with an incommensurability that grows with doping[55]. Indeed, the incommensurate ground state for $NaNiH_2$ and nearly commensurate for $LiNiH_2$ are consistent with a significant and a negligible self-doping, respectively, as found integrating the Fermi surface volume of the paramagnet.

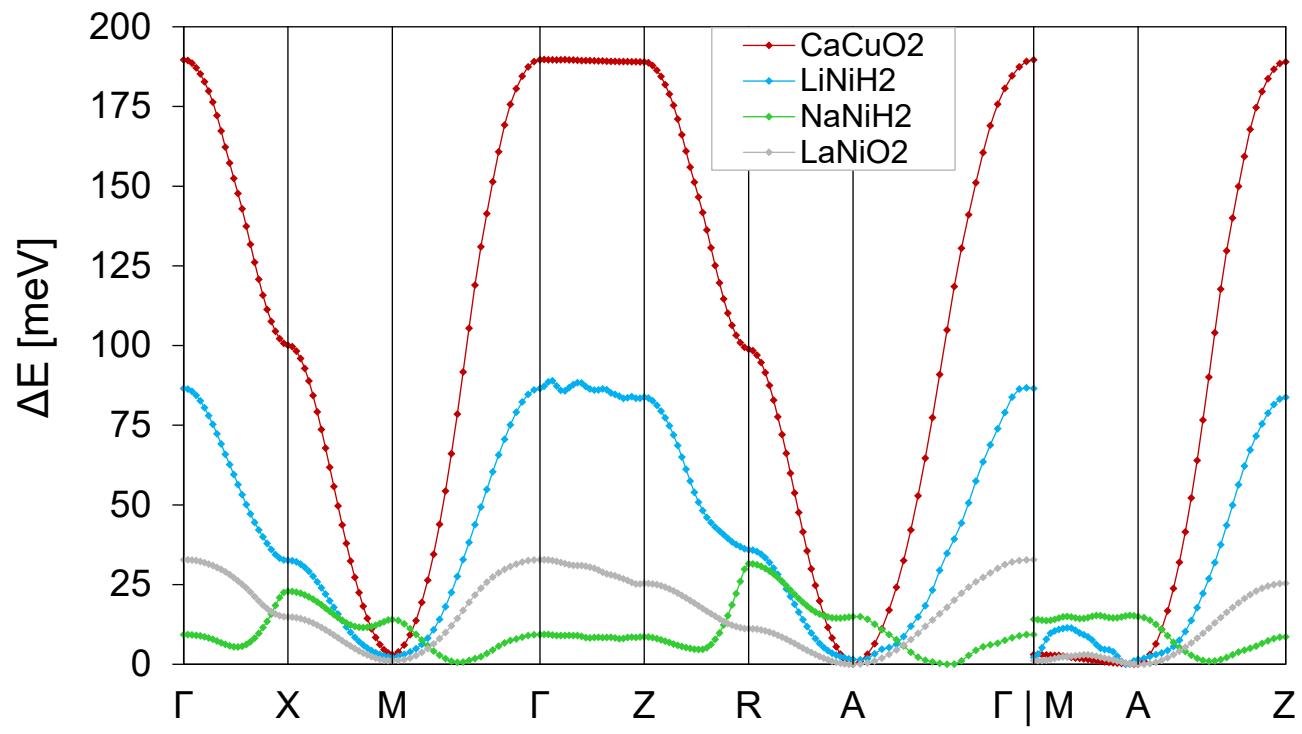

**Fig. 2.** Dispersion of spin-spiral energy for $CaCuO_2$, $LiNiH_2$, $NaNiH_2$, and $LaNiO_2$ systems calculated on GGA+U level ($U$ = 7 eV for all).

Magnetic moments evaluated in the commensurate phases (0.6-0.7 $\mu_B$, see **Table S2**[37]) are close to the value for $CaCuO_2$ (0.54 $\mu_B$) and reduced from 1 $\mu_B$ due to covalency. In contrast, for $LaNiO_2$ we find that the magnetic moment is nearly 1.0 $\mu_B$ indicating lower covalency.

The maximum $T_c$ attainable in cuprates is believed to scale with the strength of superexchange interactions $J_{2D}$[58–61]. The superexchange interaction on Ni hydrides can be estimated from the energy of spirals in the case of $LiNiH_2$ which is nearly undoped.

From the energy of spirals in **Fig. 2** we see that by choosing $U = 7$eV, $LiNiH_2$ shows an energy of magnetic interactions intermediate between cuprates and nickelates. Reducing the Hubbard parameter to the more realistic $U = 6$ eV (**Fig. S4**[37]) increases the magnetic stiffness, defined by the curvature of the ground state energy *vs.* wave vector around its minimum value, suggesting an even larger magnetic energy scale. However, $LiNiH_2$, shows a collapse to a nonmagnetic state for the highest energy spirals, indicating again a less correlated regime than cuprates. For self-doped $NaNiH_2$, instead, the energy of magnetic excitations is suppressed for both values of $U$ (**Fig. 2** and **S4**[37]) which we attribute to self-doping. This suggests that 4% doping is close to the critical doping for the suppression of magnetism due to quantum fluctuations. Local magnetic moments are expected to persist as for $U = 5$, 6 eV there is no collapse to a nonmagnetic state confirming the more correlated nature of $NaNiH_2$ compared to $LiNiH_2$.

*Superconductivity*. To predict the phase diagram in cuprate-like systems is notoriously difficult as charge, spin, and superconductivity instabilities must be considered on an equal footing and are beyond our scope. However, the leading symmetry of the order parameter in $NaNiH_2$ can be obtained by a mean-field computation in the *t-J* model based on the thigh-binding fit of the $d_{x^2-y^2}$ band. Using a Gutzwiller renormalization factor as in Ref.[62] and J=100 meV we obtained the critical temperature for d-wave and extended s-wave symmetry as shown in **Fig. 3**. Both for hole ($\delta$>0) and electron ($\delta$<0) doping the dominant order parameter is d-wave with a larger $T_c$ for hole doping as for cuprates. This suggests that the neck does not significantly affect high-$T_c$ superconductivity within a magnetic mechanism. While we expect the symmetry of the order parameter to be a robust feature of the phase diagram, $T_c$ values should be taken with a pint of salt as they are very sensitive to competition with other phases and to the value of $J$. In a more realistic computation, the superconducting dome will split in two with antiferromagnetism, not considered here, dominating in the gray-shaded region.

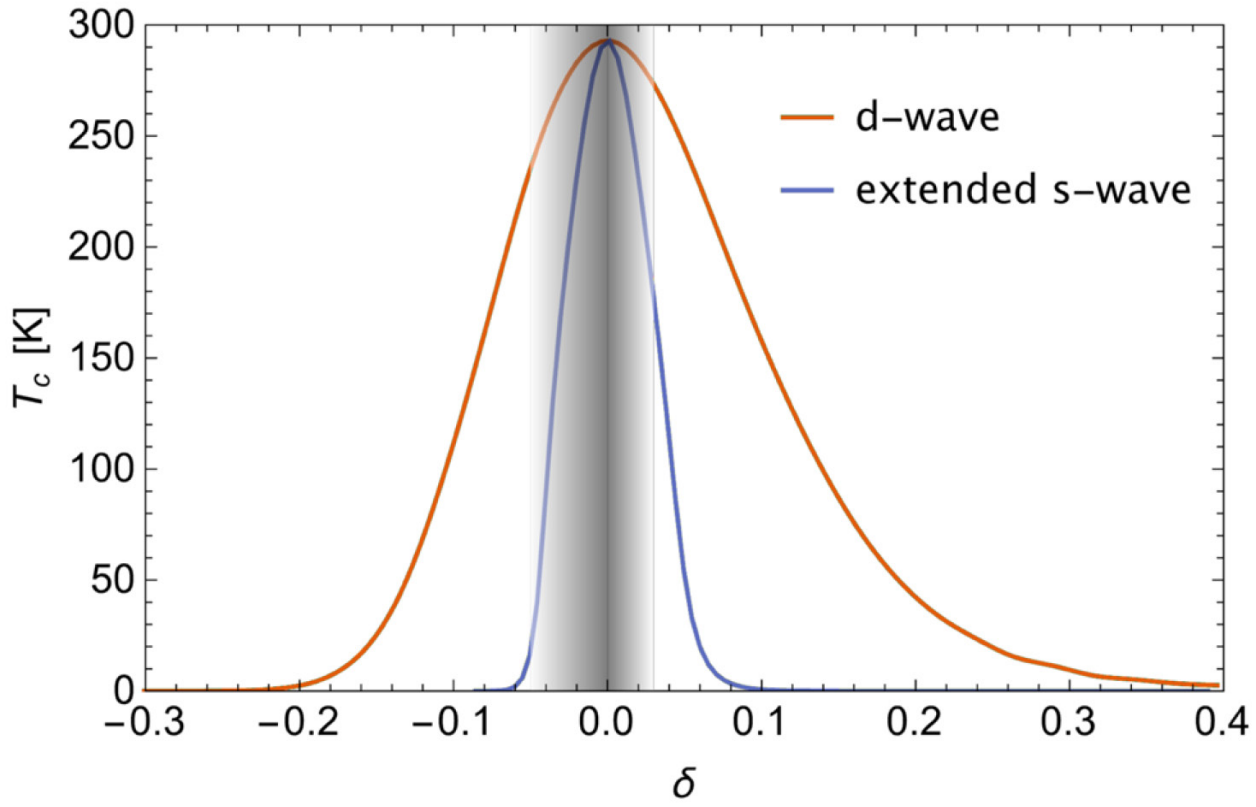

**Fig. 3.** Critical temperature for d-wave and extended s-wave superconductivity as a function of doping $\delta$ for $NaNiH_2$ within a *t-J* model computation with J=100 meV (see SI). The gray shading indicates the region in which magnetic order is expected to suppress superconducting order. Positive $\delta$ corresponds to hole doping.

*Thermodynamic stability.* Having established ternary nickel(I) hydrides to be very interesting cuprate analogs we now address the problem of their stability. We have calculated phonon energy dispersion on the GGA+U ($U$ = 6 eV) level for both $MNiH_2$ (M = Li, Na) using the PHONOPY[63,64] program. For simplicity, we considered the G-type antiferromagnetic solution, which is the collinear ground state for the chosen method. Importantly, we find dynamical stability (**Fig. S5**[37]). This ensures that if the compound is synthesized it will not transform to a different phase. We do not expect the emergence of dynamic instability with the variation of the $U$ parameter as we find dynamic stability also for nonmagnetic solutions with pure GGA (not shown).

In solids, large barriers usually protect a dynamically stable phase from transitioning into energetically more favorable phases, so the decay requires a considerable thermodynamic drive. Classical examples are the allotropes of carbon, where large energies are required to transform one into another. Also, many of the most interesting transition metal oxides are in fact metastable[65,66]. Still, synthesis of such metastable phases may be challenging. Therefore, we evaluate the thermodynamic drive of $MNiH_2$ CCO phases with respect to alternative compounds or polymorphs in terms of lattice energy using DFT methods.

To test the sensitivity of the total energy computations to the methodology we realized GGA+U calculations with different $U$ values and with a range-separated hybrid functional (HSE06[67]), as reported in **Table S2.** Although numerically more involved, hybrid functionals are generally considered more accurate for correlated insulators with open *d-shells*-, thus we focus on this method below.

The simplest route to CCO nickel hydrides could be the direct hydrogenation of the Ni and MH 1:1 molar mixture. Indeed, the closely related perovskite $LiNiH_3$ has been recently obtained *via* direct hydrogenation of Ni and LiH 1:1 molar mixture at high temperatures (3 GPa, 873 K)[68]. This result is encouraging but also presents a challenge to the synthesis of $LiNiH_2$ if there is a strong energetic drive towards $LiNiH_3$. We have found that an energetically favorable route in this direction is the decomposition $2LiNiH_2 \rightarrow LiH + Ni + LiNiH_3$, in fact, a redox reaction. As shown in **Table S2**[37] this route for the decomposition of $LiNiH_2$ is energetically favorable, with 11 meV/atom according to HSE06. A similar result is found for the analogous decomposition of $NaNiH_2$, where the energetic drive is higher, about 43 meV/atom. Simultaneously, we note the results for this reaction using GGA+U methods give similar energy differences (**Table S2**[37]). Importantly, we consider the energy gain in the redox decomposition of the $MNiH_2$ CCO phases to be too small to represent a problem for stability (see discussion below).

We also considered the decomposition of ternary nickel hydrides into binary hydrides $MNiH_2 \rightarrow NiH + MH$. In the case of both M = Li and Na, we find this decomposition routes energetically unfavorable using HSE06, by 45 meV/atom for $LiNiH_2$ and 3 meV/atom for $NaNiH_2$, while generally less accurate GGA+U methods rather suggest weak metastability of $MNiH_2$ on this path (**Table S2**[37]). This means that the opposite reaction, the formation of the desired ternary hydride $NiH + MH \rightarrow MNiH_2$, should be preferred thermodynamically.

Another question is the stability of the $MNiH_2$ CCO phases compared to other polymorphs. Especially, $MNiH_2$ can also form a NaCl type crystal structure, as shown in **Fig. S6**[37]. The latter case was considered as an intermediate phase of $LiNiH_2$ in recent experimental work[68], although not found during X-ray diffraction studies. The energetic stability calculations of $LiNiH_2$ using HSE06 indicate that the NaCl phase is favored over the CCO phase by only 7 meV/atom, while other DFT methods differ marginally suggesting possible metastability (**Table S2**[37]). High-pressure synthesis can be excluded as the enthalpic effect should further favor the NaCl phase as we predict an 11% larger volume for the CCO type. Fortunately, for $NaNiH_2$ the CCO phase is considerably preferred over the NaCl phase by 206 meV/atom with HSE06, in compliance with the GGA+U methods. This substantial stability of the $NaNiH_2$ CCO phase, over both the NaCl phase and binary hydrides, opens more possibilities for the synthesis.

To estimate the thermodynamic stability at finite temperatures we computed the vibrational contribution to the Helmholtz free energy of both $LiNiH_2$ polymorphs and the competing phases using the GGA+U ($U$ = 6 eV) method. The vibrational entropy of the CCO polymorph is relatively high, especially higher than that of the lower volume NaCl polymorph causing stabilization of the $LiNiH_2$ CCO phase with increasing temperature (**Fig. S7**[37]). We would like to emphasize here, that many known ternary inorganic compounds are metastable[66] on a similar scale – it was found that 50.5±4% of unique crystal phases in the Inorganic Crystal Structure Database are metastable and half of the identified metastable ternary compounds have DFT energy of +6.9 meV/atom or greater above ground state. Also, it was found that the standard deviation of formation energies calculated with DFT is around 24 meV/atom[69] compared to experimental values, although the study referred only to oxides, not hydrides. Thus, we predict that synthesis of CCO $LiNiH_2$ is possible, with higher probability in high temperature and moderate external pressure conditions (below GPa scale).

We stress again that the small energetic advantage of solid-state competing phases found at low temperatures is not an obstacle to synthesis. With the same methods, we investigate similar materials, namely $CaCuO_2$ and $LaNiO_2$ (**Table S2**[37]). For example, we find the decomposition $CaCuO_2 \rightarrow CuO + CaO$ slightly favorable, 1 meV/atom with HSE06. An even more striking case is $LaNiO_2$ being considerably unstable towards redox reaction by 63 meV/atom. Importantly, we find $LaNiO_2$ instability towards redox reaction at low temperatures, regardless of the DFT method used but also considerably larger than for both ternary hydrides, for which we find the maximum instabilities of 11 meV/atom ($LiNiH_2$) and 43 meV/atom ($NaNiH_2$). Therefore, since $LaNiO_2$ was successfully obtained[17–19], we expect the synthesis of both $LiNiH_2$ and $NaNiH_2$ to be chemically attainable and persist at low temperatures as metastable phases. Polymorph selection and stabilization of desired phases can be achieved through: Na/Li cationic doping, strain engineering on thin films, and using appropriate substrates as templates[6,70]. An interesting strategy could be to start from the more stable $MNiH_3$ phase and favor hydrogen desorption by an atmosphere with small hydrogen partial pressure. This will give access to hole-doped $MNiH_{2+\delta}$ phases where superconductivity is expected to be more robust with respect to the electron-doped counterpart.

*Conclusions*. We present ternary nickel hydrides as novel materials of feasible synthesis that emulate the physics of cuprates but possess unique properties including the IAE band. The charge transfer energy is notably lower than in $LaNiO_2$, which, in the context of cuprates, is associated with a higher $T_c$. The cuprate-like bands in $MNiH_2$ are broader than those found in oxides due to shorter bond lengths and greater covalency. Assuming a one-band model with a similar $U$ as in cuprates, this translates into a weaker coupling regime than cuprates but still a robust magnetic interaction. The magnetic moments found suggest that the materials are in an intermediate coupling regime as cuprates. This implies that doping on top of the intrinsic carriers, strain, or other non-thermal parameters may drive the system to a quantum-critical point and promote a superconducting phase. A *t-J* model predicts that the dominant phase should be d-wave for all doping levels. The $T_c$ from this computation is probably overestimated due to the lack of competition with other phases. Overall, based on previous work[61], and given the similarity of the band structure with cuprates and the strength of the magnetic interactions, we expect that optimally doping the $NiH_2$ planes should lead to a superconducting $T_c$ in the scale of 100 K. For the case of $NaNiH_2$, substantial self-doping opens the possibility that the stoichiometric system is already metallic and possibly superconducting.

## ACKNOWLEDGMENTS

This work was supported by the Polish National Science Centre within grant No. 2020/37/N/ST3/04206, by the Italian Ministry of University and Research through the project Quantum Transition-metalFLUOrides (QT-FLUO) PRIN 20207ZXT4Z, and by the CNR-CONICET NMES project. Computations have been carried out using resources provided by the Wroclaw Centre for Networking and Supercomputing (WCSS, http://wcss.pl), the Interdisciplinary Centre of Mathematical and Computational Modelling (ICM) at the University of Warsaw (grant GA83-34).

# Supplemental Material

## Ternary nickel hydrides: a platform for unconventional superconductivity and quantum magnetism

Mateusz Domański[1]*, Antonio Santacesaria[2], Paolo Barone[3], José Lorenzana[2]* and Wojciech Grochala[1]

(1) Center of New Technologies, University of Warsaw, Żwirki i Wigury 93, 02-089, Warsaw, Poland
(2) ISC-CNR Institute for Complex Systems and Department of Physics, Sapienza University, Piazzale Aldo Moro 5, 00185, Rome, Italy
(3) SPIN-CNR Institute for Superconducting and other Innovative Materials and Devices, Area della Ricerca di Tor Vergata, Via del Fosso del Cavaliere 100, 00133 Rome, Italy
*E-mail: m.domanski@cent.uw.edu.pl, jose.lorenzana@cnr.it

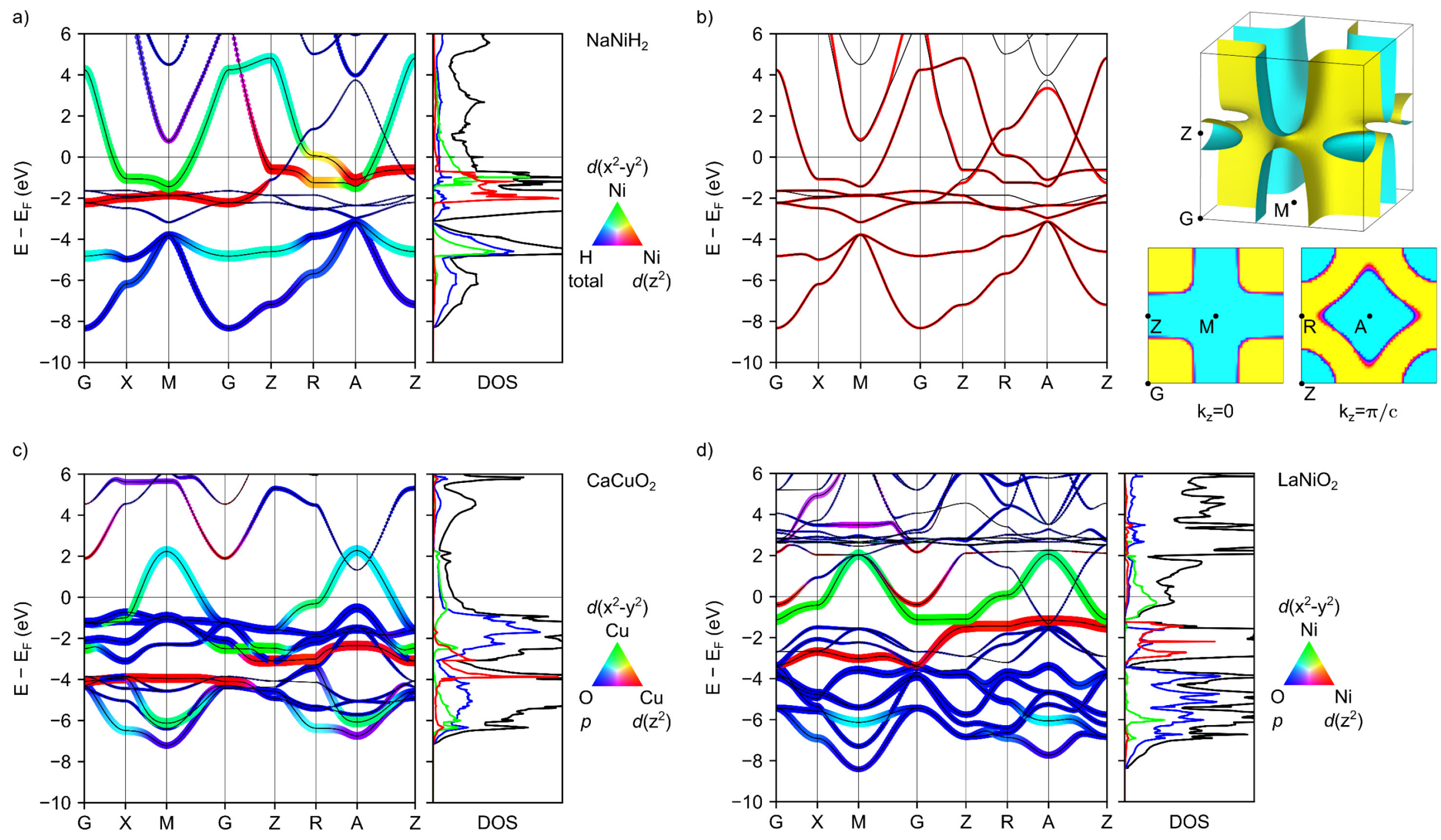

**Fig. S1.** GGA+U band structures and density of states for $NaNiH_2$ (a), $CaCuO_2$ (c), and (d) $LaNiO_2$ in CCO type structures for paramagnetic states. Black lines indicate eigenvalues of energy bands, diameters of colored circles indicate the sum of the projected states in each band, while color indicates mixing (i.e. relative ratio). The mixing of states is visualized on a Maxwell color triangle. (b) 8-band Wannier fit (red lines) of the $NaNiH_2$ GGA+U band structure (black) and obtained the Fermi surface.

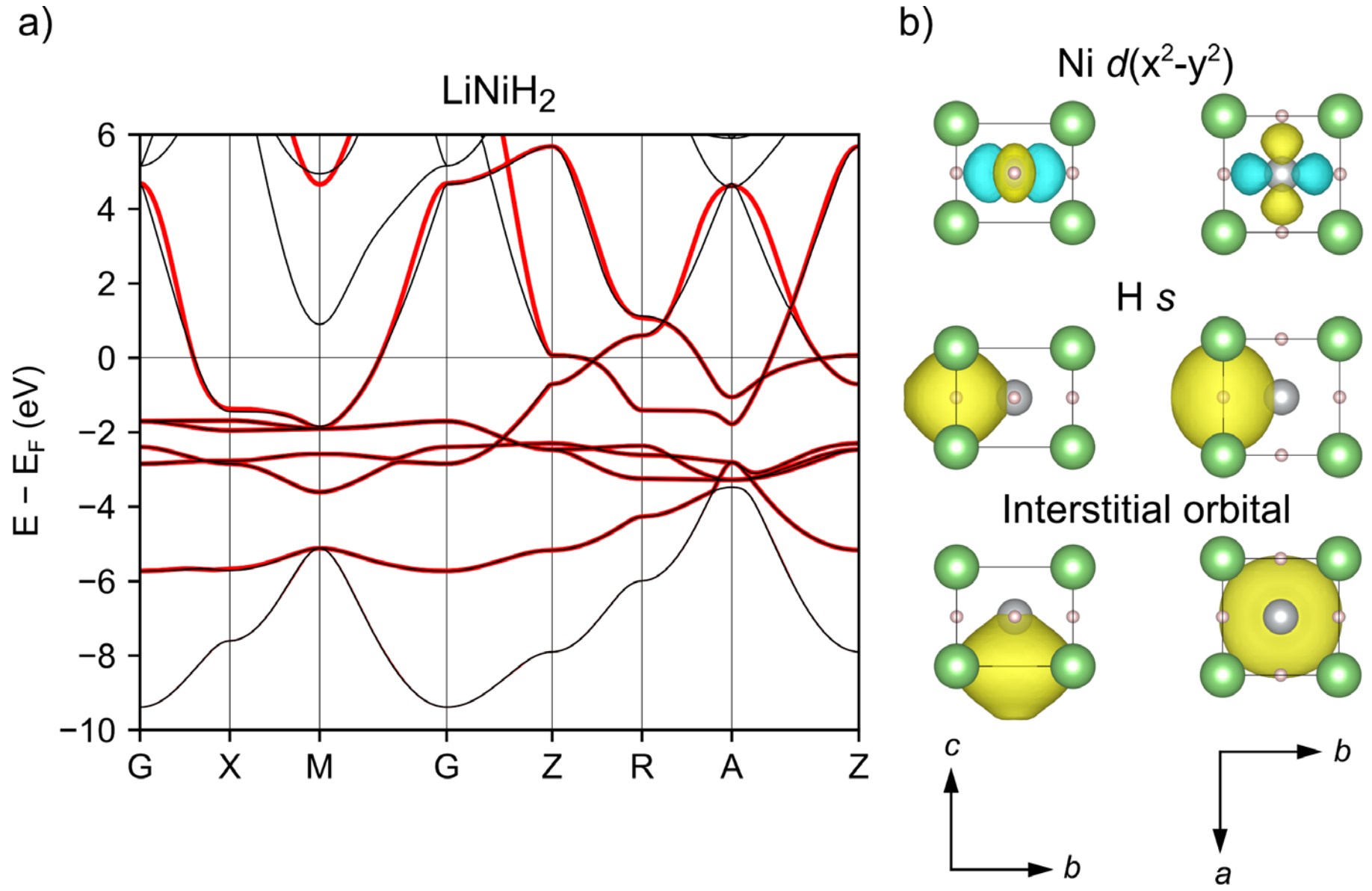

**Fig. S2.** (a) 8-band Wannier fit (red lines) of the $LiNiH_2$ GGA+U band structure (black).(b) The MLWF orbitals obtained from the Wannier fit of GGA+U bands of $LiNiH_2$. Green spheres: Li, grey spheres: Ni, small pink spheres: H.

## Tight-binding fit and t-J model computation of the superconducting order parameter phase diagram

We performed a two-band tight-binding fit representing the $d_{x2-y2}$ and the IAE bands as appropriate for $NaNiH_2$. The case of $LiNiH_2$ requires in addition the $d_{z2}$ band and will be presented elsewhere. The two bands do not hybridize significantly so we can treat them separately. We parameterize the dispersion relation of the $d_{x2-y2}$ band with the following expression:

$$(1) \quad \epsilon_d(\boldsymbol{k}) = \epsilon_d^0 - 2t(c_x + c_y) - 4t'(k_z)c_x c_y - 2t''(c_{2x}+c_{2y}) - 2t_{001}c_z - 2t_{002}c_{2z} - 2t_{003}c_{3z}$$

with $t'(k_z) = t_{110} + 2(t_{111}c_z + t_{112}c_{2z} + t_{113}c_{3z})$ and we defined $c_{nx} = \cos(n\, k_x\, a)$, etc.

For the IAE we use:

$$\epsilon_e(\boldsymbol{k}) = \epsilon_e^0 - 2t_e(c_x + c_y) - 4t_e'(k_z)c_x c_y - 2t_{ez}c_z$$

In Eq. (1), *t, t'(0)* and *t''* represent the hopping process to first, second, and third neighbors in the plane. *t'* acquires an effective dispersion due small hopping matrix element along z ($t_{001}$, $t_{002}$, $t_{003}$). These long-range hopping processes along z are needed to describe the narrow momentum neck structure.

For the mean-field superconducting computations we assumed a *t-J* model Hamiltonian using the full three-dimensional band structure but keeping only the dominant antiferromagnetic interaction in the plane. As in Ref.[61] we treated the constraint of non-double occupancy in the Gutzwiller approximation with an overall Gutzwiller renormalization factor $z = |\delta|$, where $\delta$ is the doping. Setting for simplicity $\epsilon_d^0 = 0$, at the Gutzwiller level the Hamiltonian reads,

$$H = z \sum_{\boldsymbol{k}\sigma} \epsilon_d(\boldsymbol{k})\, C_{\boldsymbol{k}\sigma}^{\dagger} C_{\boldsymbol{k}\sigma} + J \sum_{<ij>} (\boldsymbol{S}_i . \boldsymbol{S}_j - \frac{n_i n_j}{4}),$$

where $C_{\boldsymbol{k}\sigma}^{\dagger}$, $C_{\boldsymbol{k}\sigma}$ are creation and destruction operators for Gutzwiller quasiparticles with momentum $\boldsymbol{k}$ and spin $\sigma$; $\boldsymbol{S}_i$ and $n_i$ are spin and charge operators for site $i$ and $< ij >$ indicates the sum over planar nearest-neighbor bonds. For each symmetry, the critical temperature is obtained by decoupling the second term in the Hamiltonian in the Cooper channel which leads to a self-consistent gap equation. Solving the linearized gap equation for each symmetry leads $T_c$ for arbitrary doping as shown in **Fig. 3** of the main text.

**Table S1.** Tight-binding parameters for the two-band model for $NaNiH_2$**.** All values are in eV.

| $\epsilon_d^0$ | t | $t_{110}$ | t'' | $t_{001}$ | $t_{002}$ | $t_{003}$ | $t_{111}$ | $t_{112}$ | $t_{113}$ | $\epsilon_e^0$ | $t_e$ | $t_e'$ | $t_{ez}$ |
|---|---|---|---|---|---|---|---|---|---|---|---|---|---|
| 0.70 | -0.70 | -0.27 | 0.045 | 0.18 | -0.072 | 0.18 | -0.021 | 0.01 | -0.003 | 5.0 | 0.52 | 0.046 | -1.8 |

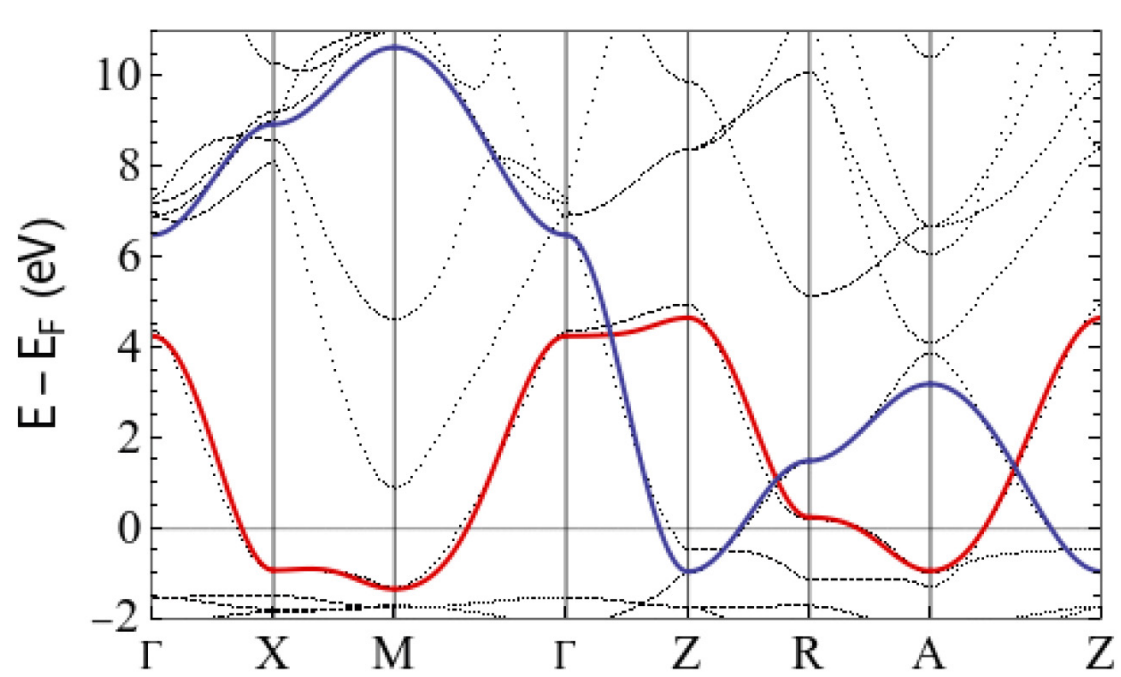

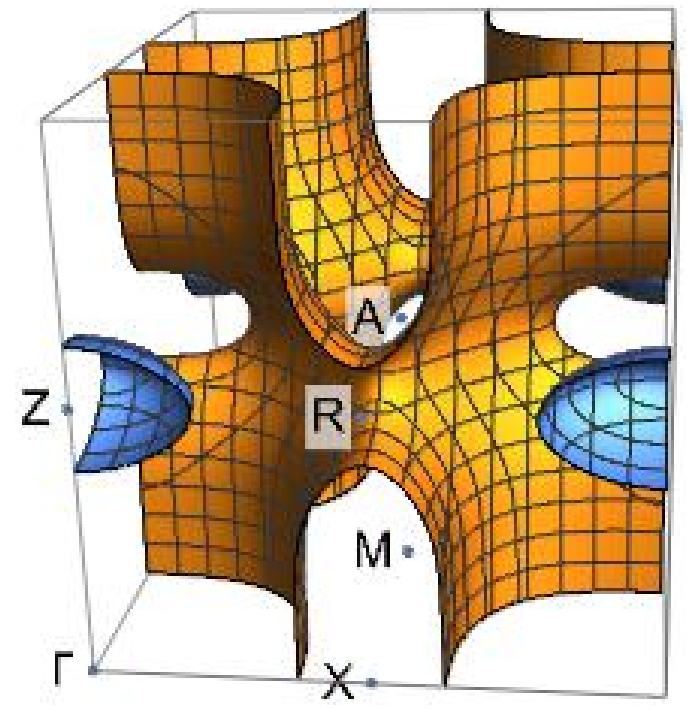

**Fig. S3.** Tight binding fit of the electronic structure of $NaNiH_2$ close to the Fermi level and Fermi surface. The $d_{x2-y2}$ band is shown in red (left) and the Fermi surface is shown in orange-yellow (right). The IAE features are shown in blue on both panels.

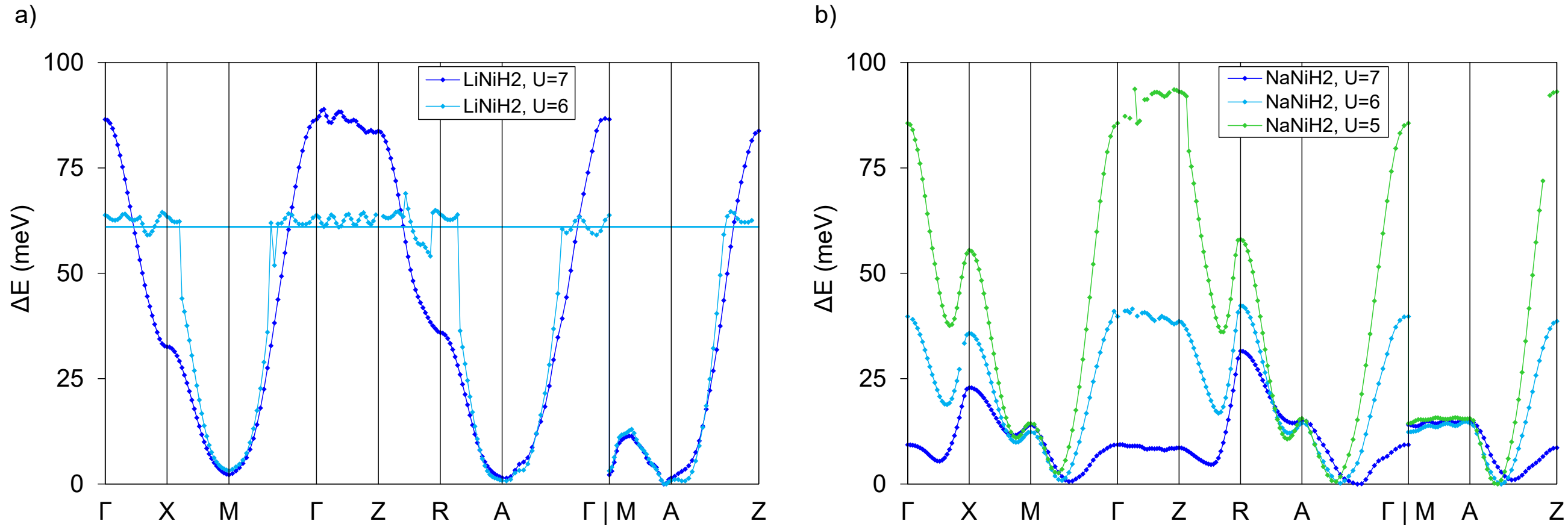

**Fig. S4.** Dispersion of spin-spiral energy for $LiNiH_2$ and $NaNiH_2$ with lower *U* values. (a) Dispersion of spin-spiral energy (meV per Ni) for $LiNiH_2$ system calculated on GGA+U level with *U* = 6 and 7 eV. The horizontal line refers to the energy of paramagnetic state for *U* = 6 eV showing that the spiral solution collapses to the paramagnetic solution at the energy crossing. For $LiNiH_2$ with *U* = 5 eV we notice lack of convergence of spin-spiral calculations due to the fact that paramagnetic state is only +4.8 meV/Ni higher than collinear magnetic groundstate (AF-C). (b) Dispersion of spin-spiral energy for $NaNiH_2$ system with increasing U (5, 6, 7 eV) calculated on GGA+U level. Values of *U* in the labels are provided in eV.

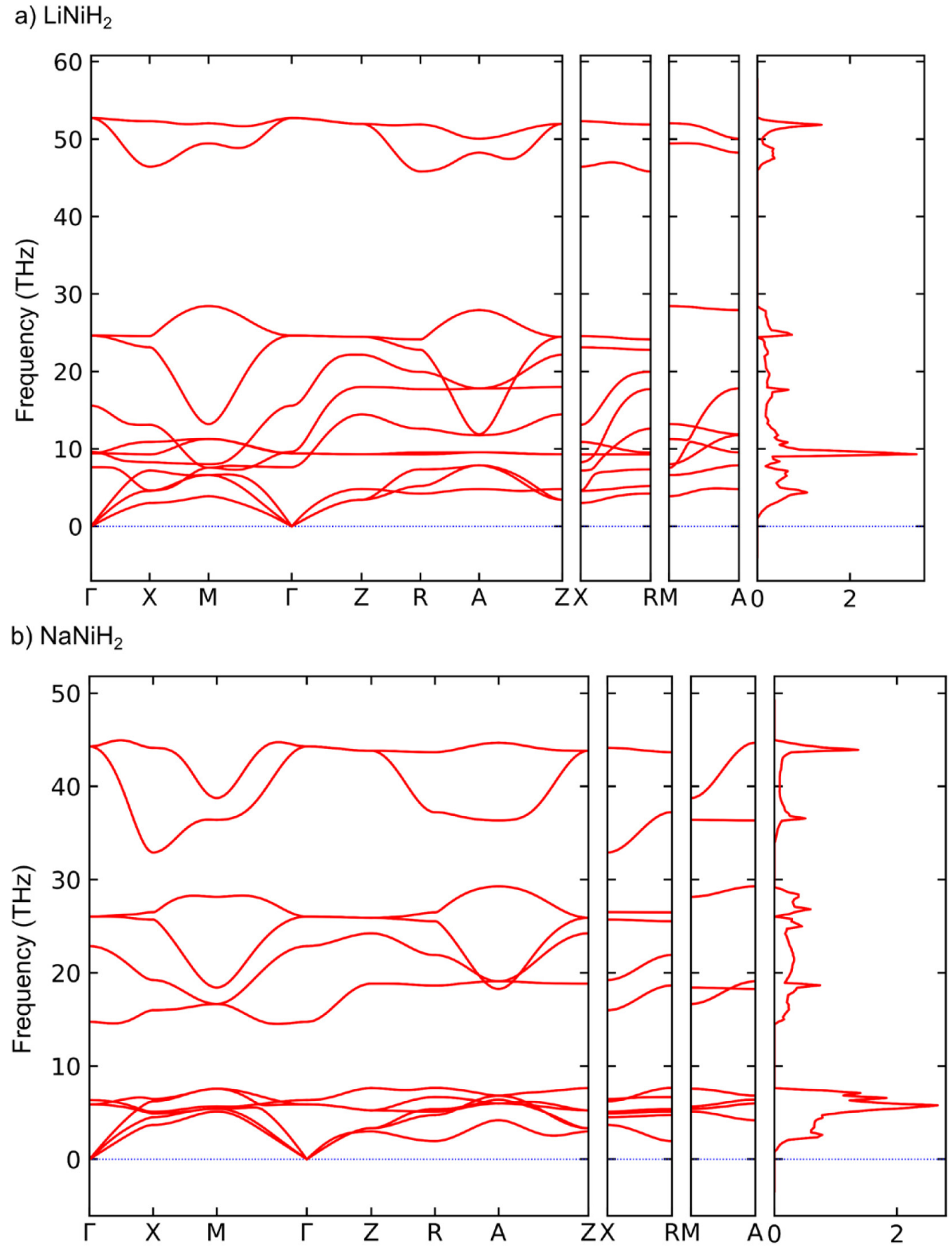

**Fig. S5.** Phonon energy dispersion and density of states for $LiNiH_2$ (AF G-type) and $NaNiH_2$ (AF C-type) systems in CCO structure calculated on GGA+U level (*U* = 6 eV for both). Calculations were perform with the PHONOPY[62,63] program with the finite displacement method. Calculations of phonon Γ-point frequencies, dispersion, and DOS were carried out for the 2×2×2 supercells. For phonon calculations denser k-spacing of 0.08 Å$^{-1}$ was used with Gaussian smearing of 0.05 eV.

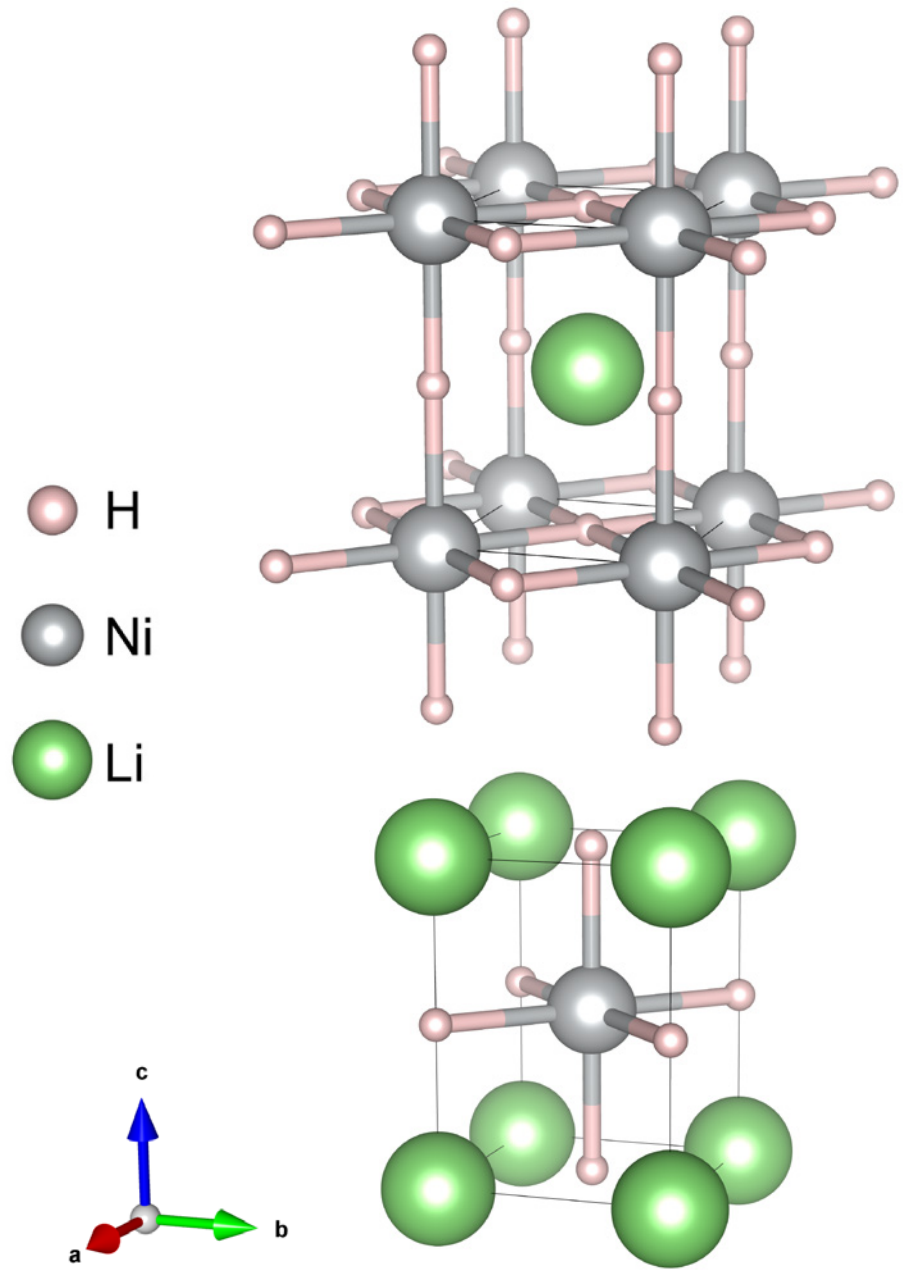

**Fig. S6.** The NaCl polymorph of $LiNiH_2$. In this phase $Ni^+$ cations are coordinated octahedrally having an axial compression along *c* lattice vector, meaning $d^9$ localizes on $d(z^2)$ orbital.

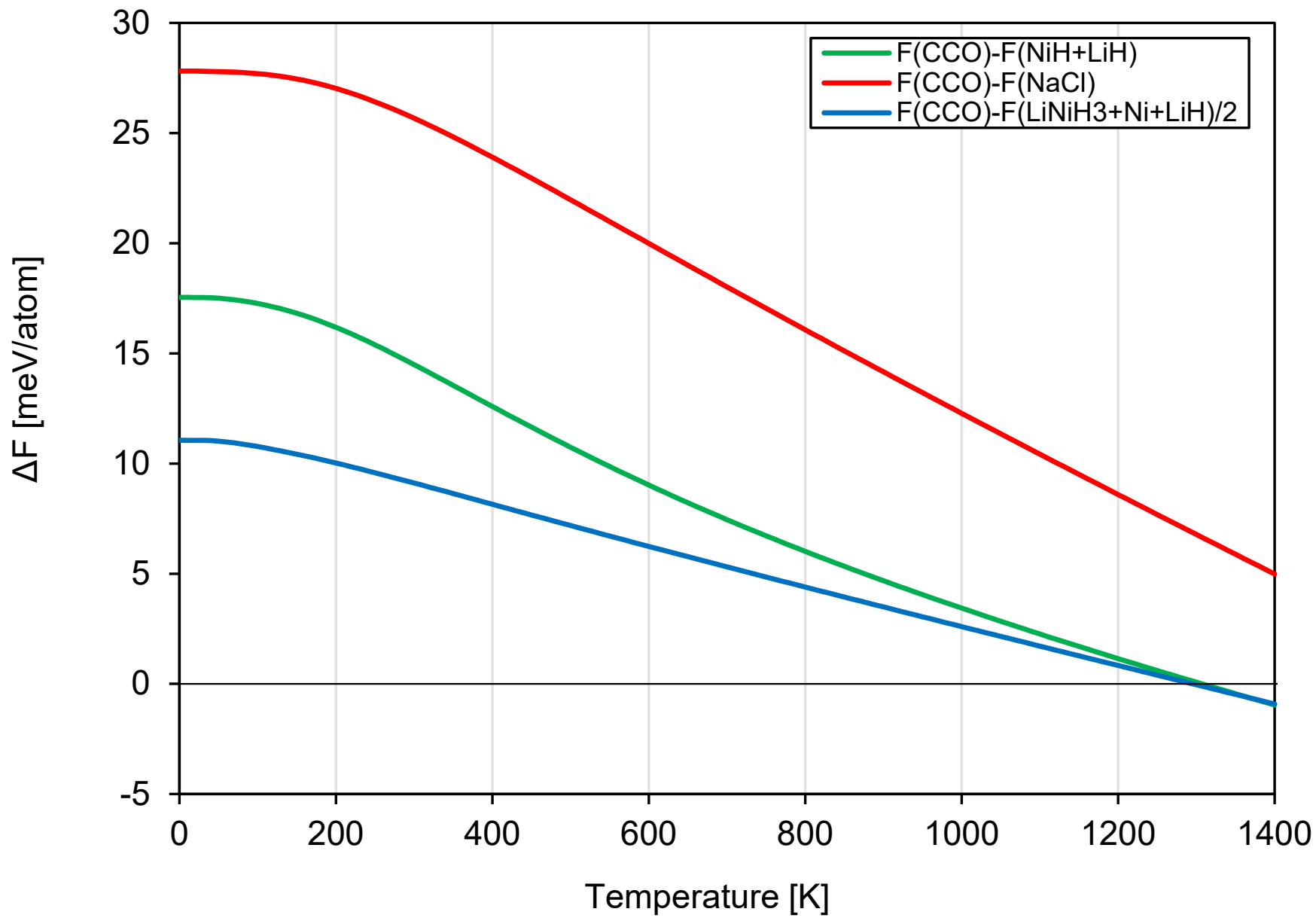

**Fig. S7.** Difference in Helmholtz free energy (ΔF) of $LiNiH_2$ CCO phase versus competing phases, calculated with GGA+U (*U* = 6 eV for Ni).

**Table S2.** Energetic stability of $MNiH_2$ (M = Li, Na) CCO phases calculated with GGA+U (*U* in eV) and a range-separated hybrid functional (HSE06[66]). For reference, we compute analogous energetic stability for $LaNiO_2$ and $CaCuO_2$. Energetic stability is calculated as the difference in lattice energies in the CCO formation reactions given in each column header, including formation from NaCl polymorph. Electronic energy values are given in eV/atom in the formula. Negative values (gray shading) indicate the stability of the ternary CCO phases. Additionally, we provide absolute magnetizations (in $\mu_B$) of Ni/Cu atoms in the CCO ternary phases for their collinear groundstates. The hybrid functional introduces a portion of Hartree-Fock non-local exchange, solving problems with excessive electron delocalization seen for semilocal GGA functionals.

| ***LiNiH₂*** | Method | U (eV) | Groundstate | $M(Ni^{+})$ | ΔE (eV) $NiH+LiH \rightarrow LiNiH_2$ | ΔE (eV) $\frac{1}{2}\cdot(LiNiH_3+Ni+LiH) \rightarrow LiNiH_2$ | ΔE (eV) $NaCl \rightarrow CCO$ |
|---|---|---|---|---|---|---|---|
| | GGA | 0 | PM | 0 | 0.025 | 0.039 | 0.017 |
| | | 5 | AF-C | 0.301 | 0.015 | 0.021 | 0.015 |
| | | 6 | AF-G | 0.551 | 0.006 | 0.011 | 0.027 |
| | | 7 | AF-G | 0.629 | -0.010 | -0.007 | 0.041 |
| | HSE06 | - | AF-C | 0.682 | -0.045 | 0.011 | 0.007 |

| ***NaNiH₂*** | Method | U (eV) | Groundstate | $M(Ni^{+})$ | ΔE (eV) $NiH+NaH \rightarrow NaNiH_2$ | ΔE (eV) $\frac{1}{2}\cdot(NaNiH_3+Ni+NaH) \rightarrow NaNiH_2$ | ΔE (eV) $NaCl \rightarrow CCO$ |
|---|---|---|---|---|---|---|---|
| | GGA | 0 | PM | 0 | 0.102 | 0.110 | -0.229 |
| | | 5 | AF-G | 0.549 | 0.073 | 0.080 | -0.193 |
| | | 6 | AF-C | 0.687 | 0.053 | 0.064 | -0.170 |
| | | 7 | AF-C | 0.754 | 0.029 | 0.043 | -0.148 |
| | HSE06 | | AF-C | 0.679 | -0.003 | 0.043 | -0.206 |

| ***LaNiO₂*** | Method | U (eV) | Groundstate | $M(Ni^{+})$ | ΔE (eV) $\frac{1}{2}\cdot(La_2O_3+NiO+Ni) \rightarrow LaNiO_2$ |
|---|---|---|---|---|---|
| | GGA | 6 | AF-G | 0.996 | 0.141 |
| | | 7 | AF-G | 1.026 | 0.050 |
| | HSE06 | - | AF-C | 0.966 | 0.063 |

| ***CaCuO₂*** | Method | U (eV) | Groundstate | $M(Cu^{2+})$ | ΔE (eV) $CaO+CuO \rightarrow CaCuO_2$ |
|---|---|---|---|---|---|
| | GGA | 7 | AF-G | 0.541 | 0.005 |
| | | 9 | AF-G | 0.608 | 0.016 |
| | HSE06 | - | AF-G | 0.565 | 0.001 |